\documentclass[a4paper,fleqn,usenatbib]{mnras}

\usepackage[T1]{fontenc}
\usepackage{ae,aecompl}

\usepackage[pdftex]{graphicx}	
\usepackage{epstopdf}
\usepackage{amsmath}	
\usepackage{amssymb}	
\usepackage{xspace}
\usepackage{textcomp}
\usepackage{gensymb}
\usepackage{enumerate}

\usepackage{ifthen}
\def\draftversion{0} 

\makeatletter
\newcommand\mytoc{%
    \@starttoc{toc}%
}
\makeatother

\ifthenelse{\equal{\draftversion}{0}}{
	\usepackage{xcolor}
	\newcommand{\tmp}{}
	\newenvironment{envcomm}[1]{\renewcommand{\tmp}{#1}\begin{color}{blue}\begin{center}\hrule\vspace{0.5mm}\tmp's COMMENTS\end{center}}{\begin{center}END OF \tmp's COMMENTS\vspace{0.5mm}\hrule\end{center}\end{color}}
	\newenvironment{draft}{\begin{color}[rgb]{0,0.4,0}\begin{center}\hrule\vspace{0.5mm}DRAFT\end{center}}{\begin{center}END OF DRAFT\vspace{0.5mm}\hrule\end{center}\end{color}}
	\newcommand{\comcomm}[2]{\begin{color}{blue}\ $\bullet$ \textbf{#1:} #2 $\bullet$\ \end{color}}
	\newcommand{\revend}[1]{\par\begin{color}[rgb]{0,0.4,0}\begin{center}\hrule\vspace{0.5mm}END OF #1's REVISIONS\vspace{0.5mm}\hrule\end{center}\end{color}\par}
	\newcommand{\todo}[1]{\begin{color}{red}\ $\bullet$ \textbf{To do: }#1 $\bullet$\ \end{color}}

	\newcommand{\del}[1]{\begin{color}[rgb]{0,0.5,0.0}\ $\bullet$ \textbf{Deleted: }#1 $\bullet$\ \end{color}}
	\newcommand{\sk}[1]{\begin{color}[rgb]{0.6,0,0.6}#1\end{color}}
	\newcommand{\toc}{\par\begin{color}[rgb]{0.6,0,0.6}\begin{center}\hrule\vspace{0.5mm}\begingroup\small\let\cleardoublepage\relax\let\clearpage\relax\mytoc\endgroup\vspace{0.5mm}\hrule\end{center}\end{color}\par}
	}{
	\newsavebox{\trashcan}
	\newenvironment{envcomm}[1]{\begin{lrbox}{\trashcan}\begin{minipage}{\columnwidth}}{\end{minipage}\end{lrbox}}
	
	\newcommand{\comcomm}[2]{}
	\newcommand{\revend}[1]{}
	\newcommand{\todo}[1]{}
	
	\newcommand{\del}[1]{}
	\newcommand{\sk}[1]{}
	\newcommand{\toc}{}
	}

\long\def\symbolfootnote[#1]#2{\begingroup%
\def\thefootnote{\fnsymbol{footnote}}\footnote[#1]{#2}\endgroup} 

\newcommand{\bb}[1]{\ifmmode \mbox{\boldmath $ #1$} \else  \mbox{\boldmath $#1$} \fi}

\newcommand{\U}[1]{\ensuremath{\mathrm{~#1}}}
\newcommand{\yr}{\U{yr}}
\newcommand{\Myr}{\U{Myr}}
\newcommand{\Gyr}{\U{Gyr}}
\newcommand{\pc}{\U{pc}}
\newcommand{\kpc}{\U{kpc}}

\newcommand{\msun}{\U{M}_{\odot}}
\newcommand{\Msun}{\msun}
\newcommand{\Msunyr}{\Msun\yr^{-1}}
\newcommand{\cc}{\U{cm^{-3}}}
\newcommand{\mpc}{\U{M_{\odot}\ pc^{-3}}}
\newcommand{\mpcc}{\U{M_{\odot}\ pc^{-2}}}

\newcommand{\ramses}{\texttt{RAMSES}\xspace}
\newcommand{\pygme}{\texttt{PyGME}\xspace}

\title[Formation of Nuclear Cluster]{New insights on the formation of nuclear star clusters}

\author[Guillard, Emsellem, Renaud]{
Nicolas Guillard$^{1,2}$\thanks{E-mail: nguillar@eso.org}, 
Eric Emsellem$^{2,3}$ \&
Florent Renaud$^{4}$
\\
$^{1}$Excellence Cluster Universe, Boltzmannstr. 2, D-85748 Garching, Germany\\
$^{2}$European Southern Observatory, Karl-Schwarzschild-str. 2, D-85748 Garching, Germany\\
$^{3}$Universit\'e de Lyon 1, CRAL, Observatoire de Lyon, 9 av. Charles Andr\'e, F-69230 Saint-Genis Laval; CNRS, UMR 5574; ENS de Lyon, France\\
$^{4}$Department of Physics, University of Surrey, Guildford, GU2 7XH, UK
}

\date{Accepted 2016 June 27. Received 2016 June 17; in original form 2016 May 24}

\pubyear{2016}

\begin{document}
\label{firstpage}
\pagerange{\pageref{firstpage}--\pageref{lastpage}}
\maketitle

\begin{abstract}
Nuclear Clusters (NCs) are common stellar systems in the centres of galaxies. Yet, the physical mechanisms involved in their formation are still debated. Using a parsec-resolution hydrodynamical simulation of a dwarf galaxy, we propose an updated formation scenario for NCs. In this ``wet migration scenario'', a massive star cluster forms in the gas-rich disc, keeping a gas reservoir, and growing further while it migrates to the centre via a combination of interactions with other substructures and dynamical friction. A wet merger with another dense cluster and its own gas reservoir can occur, although this is not a pre-requisite for the actual formation of the NC. The merging process does significantly alter the properties of the NC (mass, morphology, star formation history), also quenching the on-going local star formation activity, thus leading to interesting observational diagnostics for the physical origin of NCs. A population of lower mass clusters co-exist during the simulation, but these are either destroyed via tidal forces, or have high angular momentum preventing them to interact with the NC and contribute to its growth. The proposed updated scenario emphasises the role of gas reservoirs associated with the densest star clusters formed in a gas-rich low-mass galaxy. 
\end{abstract}




\section{Introduction}
\label{intro}

Nuclear Star Clusters (NCs) are present in a wide variety of galaxies, from early \citep[e.g.][]{Carollo1998, Turner2012, denBrok2014} to late-type galaxies \citep[e.g.][]{Boker2002,Georgiev2014, Carson2015}. Observational studies with the Hubble Space Telescope show that about 75\% of spiral and dwarf elliptical galaxies have a prominent NC \citep{Cote2006, Seth2006,Seth2008b,Neumayer2012}. NCs have typical sizes of a few to a few tens of parsecs and a mass  from  $10^4 \Msun$ to $10^8 \Msun$ \citep[e.g.][]{Georgiev2014}, which rank them among the densest stellar objects in the Universe. The mass of NCs roughly scales with the galactic host properties such as the galactic mass, the velocity dispersion of the spheroidal component or and the total galactic luminosity \citep[e.g.][]{Ferrarese2006, Rossa2006, Graham2012, Scott2013, Georgiev2016}. Understanding the physical origins of the properties of NC could thus shed new lights on the galaxy evolution.

\begin{figure*}
	\centering
	\includegraphics[scale=0.7]{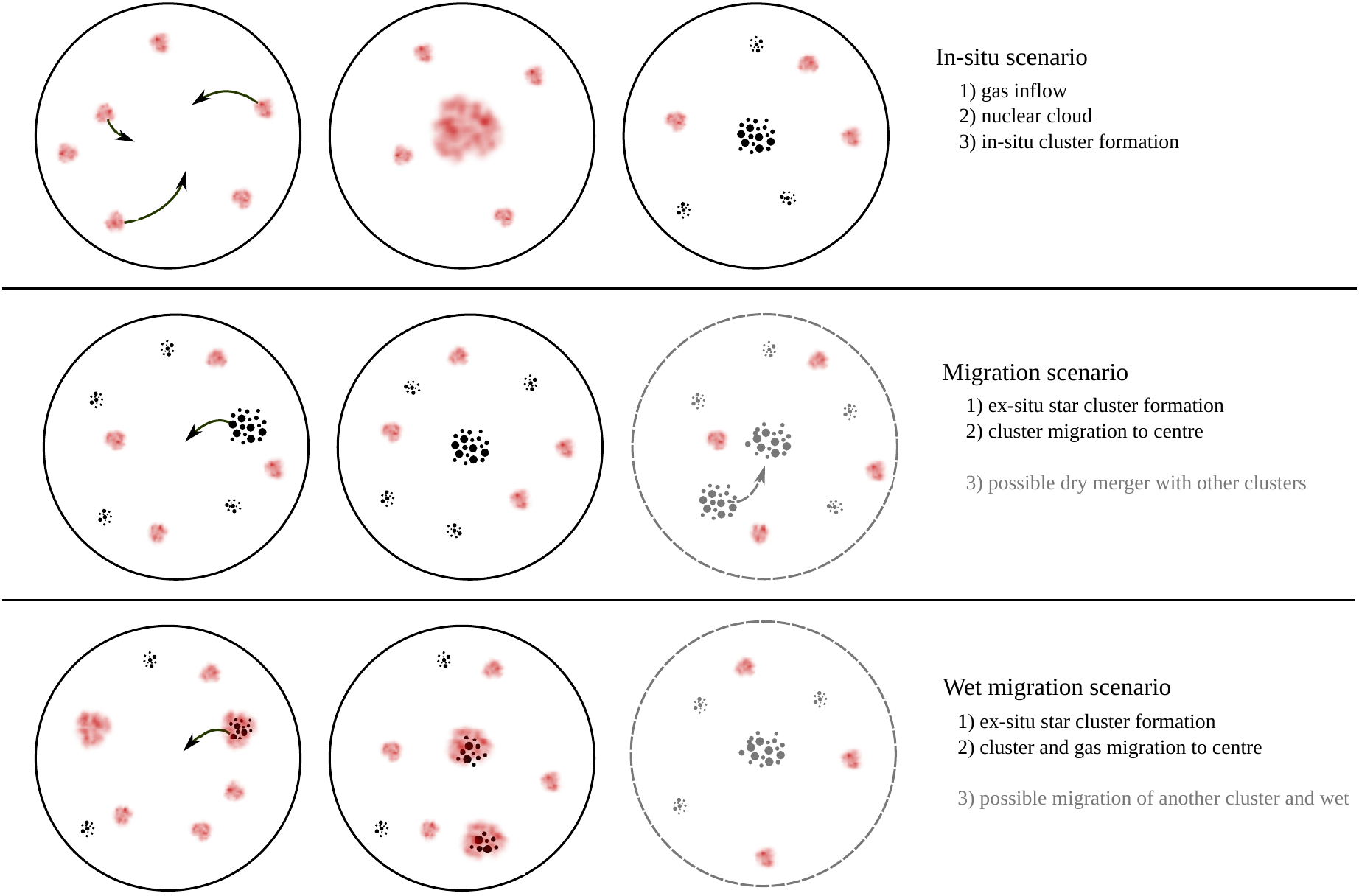}
   \caption{Schematics representation of NC scenarios from the literature and the one proposed in this work.}
  \label{fig:scenario_schm}
\end{figure*}

To date, two main formation scenarios have been proposed (see top and middle rows of Fig.~\ref{fig:scenario_schm}):
\begin{itemize}
 \item in-situ \citep{Milosav2004}: gas falls onto the galactic centre which subsequently triggers star formation in the central few parsecs and forms the NC.
 \item migration \citep{Tremaine1975}: a massive cluster forms, then migrates towards the centre by dynamical friction. This process is potentially followed by dry mergers (i.e., gas free) with other clusters \citep{Andersen2008, 2013ApJ...763...62A}. 
\end{itemize}

These formation scenarios imprint specific signatures on the properties of NCs. Probing the galaxy properties or examining the above-mentioned scaling relations should thus help us to disentangle between the various formation scenarios. The power-law relation between the mass of the NC and the velocity dispersion of the galaxy host observed by \citet{Ferrarese2006} is not reproduced by the in-situ scenario (see analytical model \citealt{2013ApJ...763...62A}), while predictions from the migration model, including a dry-merger step, seem to be more successful \citep{2013ApJ...763...62A, Arcasedda2014}. Dynamical simulations from \citet{Hartmann2011} also show that the mergers of star clusters in models tuned for NGC 4244 and M33 retrieve the properties expected from the scaling relations. 

More recent studies emphasise the fact that these two scenarios are not exclusive and likely contribute together to build the properties of the NC \citep[e.g.][]{denBrok2014, 2016arXiv160502881C}. \citet{Hartmann2011} points out that despite the fact that properties induced by cluster mergers are in agreement with observations, in-situ star formation would still contribute for $\sim50\%$ of the mass of the NC. Semi-analytic models by \citet{Antonini2015} lead to similar conclusions showing that stars formed in-situ contribute to a large fraction (up to 80\%) of the total NC mass. 

To further investigate the diverse origins of stellar population in NCs, we present a self-consistent hydrodynamical model of NCs formation in its galactic context.
Using a parsec-resolution hydrodynamical simulations of a gas-rich dwarf galaxy, we propose a new scenario for the formation of NC (see bottom panel of Fig.~\ref{fig:scenario_schm}) based on ex-situ formation of massive clusters, their continuous growth, migration to the galactic centre, potentially followed by a \emph{wet} merger with other clusters bringing their own gas reservoirs. 

In Section~\ref{num_methd}, we describe the numerical methods. The formation scenario of the NC is described in Section~\ref{story_explanation}. We show the interaction between the NC and the galactic cluster population in Section~\ref{clst_small} and finally discuss the implications of this new scenario in Section~\ref{discussion}.




\section{Numerical tools and convergence}
\label{num_methd}

We run hydrodynamical simulations of an isolated dwarf galaxy using the Adaptive Mesh Refinement (AMR) code \ramses \citep{Teyssier2002}. We define 3 types of particles: the dark matter (DM), the stars included in the initial conditions (hereafter referred to as primitive stars), and stars we formed during the course of the simulations (hereafter referred to as new stars).  The code solves the equations of motion with a particle-mesh scheme. The code uses a softening of the gravitational acceleration for the DM and primitive stars of 7~pc, while the softening for the new stars is the local resolution of the AMR grid, which is specific to each simulation (see Table~\ref{IC_table} in Sect.~\ref{num_methd_ic}). For the gas, the code solves the Euler equations on the AMR grid, allowing the densest regions to be refined while keeping a low resolution on more diffuse media. To avoid the artificial fragmentation of the densest regions, we add a pressure floor that ensures that a thermal Jeans length is always resolved by at least four cells. The physical ingredients we use in this simulation are similar to the ones used in \citet{Renaud2015}.

The size of the simulated volume is of $30 \times 30 \times 30$~kpc$^3$, with the least resolved cells spanning 120pc. We run a set of 3 simulations in which we vary the maximal resolution from 15~pc$^3$ to 3.5~pc$^3$ (see Table~\ref{IC_table}). The galaxy is modeled in isolation, thus neglecting the cosmological context. The simulations have been run on the C2PAP facilities (Excellence Cluster, Garching) for about 1 million CPU-hours on 512 cores.

The gas is heated by ultraviolet radiation and cooled down by atomic cooling tabulated at solar metallicity \citep{Courty2004}. The minimal temperature reached is of 200K. 

Star formation follows the Schmidt law: $\rho_{SFR} = \epsilon \rho /$ t\textsubscript{ff} $\propto \epsilon \rho^{3/2}$ where $\rho$ is the gas density, $\epsilon$ is the dimensionless efficiency of the star formation and t\textsubscript{ff}= $\sqrt{3\pi/(32 G \rho)} $ is the free-fall time. This only concerns densities higher than a given threshold. We set an efficiency of 2\% and a density threshold of 100~cm$^{-3}$, so that the star formation rate (SFR) of the dwarf is about $0.1 \Msun$. This corresponds to the rates observed for galaxies of $\sim10^9\Msun$ at redshift z=2-3 which is the type of galaxies we model in this work \citep[e.g.,][]{Behroozi2013}. The stellar particles have a mass of $130 \Msun$.

The stellar feedback recipes we used are described in \citet{Renaud2013}. Photo-ionization is modeled by creating a Str\"{o}mgren sphere around massive stars (20\% of the stars mass explode as SNe) younger than 10~Myr. The radius of the sphere depends on the ambient gas density and the time-varying stellar luminosity. The interstellar medium (ISM) in the sphere is heated up to $4\times10^4$~K. In the bubble, the code injects momentum-driven feedback in the form of radial velocity kicks to model radiative pressure. Type II supernova (SN) feedback is implemented as a Sedov blast wave (see  \citealt{Dubois2008} for details). SN injects $10^{51}$~erg in a kinetic form. Feedback from a potential active galactic nucleus is not included in these simulations.



\subsection{Initial conditions and final state}
\label{num_methd_ic}

Galaxies with stellar mass of $\sim 10^9 - 10^{10} \Msun$ have the highest fraction of nucleated galaxies \citep{Pfeffer2014}, and we therefore set the total baryonic mass of our galaxy model in this range, namely to $3.3\,10^9 \Msun$. Taking conditions representative of redshift $z \sim 2-3$ low-luminosity galaxies, we set the gas mass fraction to 70\% of the baryonic mass \citep{Daddi2010}, the stellar and gaseous masses being $10^9$ and $2.3\,10^9 \Msun$, respectively. The DM halo has a mass of $10^{11} \Msun$, following the scaling relation between DM halo and stellar disc from \citet[]{Ferrero2012}. We model the DM halo with a Navarro-Frenk-White (NFW) profile \citep{NFW1996} that has a concentration of 16 and a virial radius of 120~kpc. We truncate the halo at a radius of 15~kpc since we focus on the central regions of the galaxy. 

\begin{figure}
	\centering
	\includegraphics[scale=0.4]{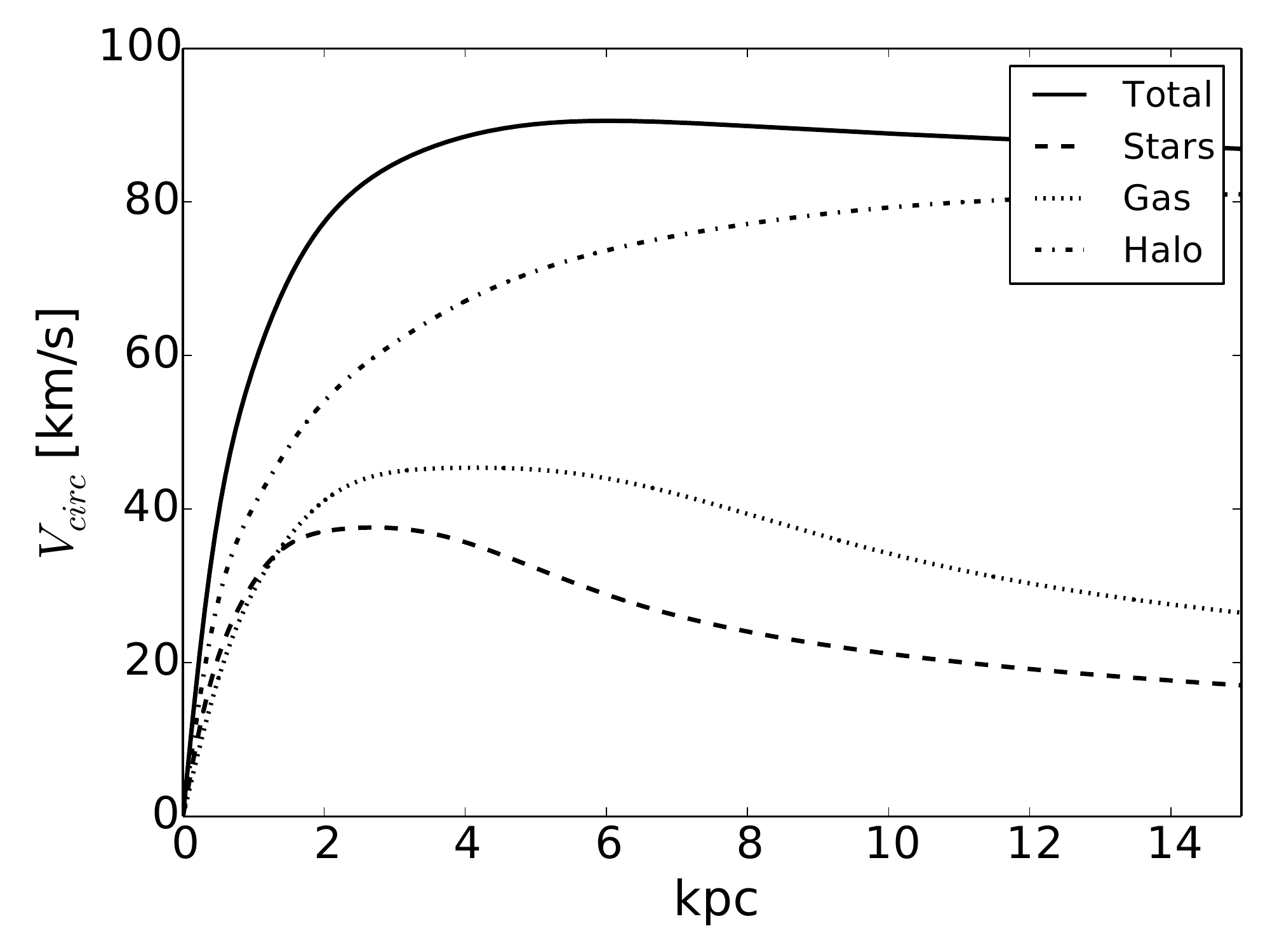}
    \caption{Rotation curves of the galactic components at $t=0$ Myr.}
    \label{fig:vcirc}
\end{figure}

At $t=0$, our simulation volume is composed of both gaseous and stellar exponential discs embedded in a dark matter halo. We use the code \pygme (Python Multiple Gaussian Expansion) to generate the stellar component, the DM and gas. This code makes use of the Multi-Gaussian Expansion method \citep{Monnet1992, Emsellem1994}, and spatially decomposes the mass of the galaxy in a set of Gaussian functions. We used a total of 26 Gaussians to generate the galaxy components: 8 for the DM Halo, 9 for the stellar disc and 9 for the gas disc. The velocities of the particles are derived via the Jeans equations considering all components (gas, stars, dark matter) for the gravitational potential. The gas particles are then replaced by AMR cells. The initial properties of the galaxy are summarized in Table~\ref{IC_table}, and Fig.~\ref{fig:vcirc} displays the initial rotation profiles of the galaxy and of its components.

\begin{table}
	\centering
	\caption{Initial conditions}
	\begin{tabular}{l|c|c|r} 
		\hline \hline
		Box length ($\kpc$) & \multicolumn{3}{|c|}{30}\\
		AMR coarse level & \multicolumn{3}{|c|}{8}\\
		AMR finest level & 11 & 12 & 13 \\
		Highest resolution (pc) & 14.6 & 7.3 & 3.7 \\
		\hline
		\textbf{DM Halo} & & &\\
		Virial mass ($\times 10^9 \Msun$) & \multicolumn{3}{|c|}{100}\\
		Virial radius ($\kpc$)  & \multicolumn{3}{|c|}{120}\\
		Cut radius ($\kpc$)  & \multicolumn{3}{|c|}{15}\\
		Concentration & \multicolumn{3}{|c|}{16}\\
		Profile & \multicolumn{3}{|c|}{Navarro-Frenk-White} \\
		Number of particles (x 10\textsuperscript{5}) & \multicolumn{3}{|c|}{37.5} \\
		
		\hline
		\textbf{Primitive stars} & & &\\
		Mass ($\times 10^9 \Msun$) & \multicolumn{3}{|c|}{1}\\
		Profile & \multicolumn{3}{|c|}{Exponential} \\
		Scale radius ($\kpc$) & \multicolumn{3}{|c|}{1} \\
		Cut radius ($\kpc$) & \multicolumn{3}{|c|}{7.5} \\
		Scale height ($\pc$) & \multicolumn{3}{|c|}{250} \\
		Cut height ($\pc$) & \multicolumn{3}{|c|}{750} \\
		Number of particles (x 10\textsuperscript{5}) & \multicolumn{3}{|c|}{15} \\
		\hline
		\textbf{Gas} & & &\\
		Mass ($\times 10^9 \Msun$) & \multicolumn{3}{|c|}{2.3}\\
		Profile & \multicolumn{3}{|c|}{Exponential}\\
		Scale radius ($\kpc$) & \multicolumn{3}{|c|}{1.65} \\
		Cut radius ($\kpc$) & \multicolumn{3}{|c|}{7.5} \\
		Scale height ($\pc$) & \multicolumn{3}{|c|}{165} \\
		Cut height ($\pc$) & \multicolumn{3}{|c|}{750} \\
		Average number of cells ($\times 10^6$) & \multicolumn{3}{|c|}{2.4}\\
		\hline \hline
	\end{tabular}
	
	\label{IC_table}
\end{table}

Star formation and feedback are not active at the beginning of the simulations. We progressively increase the refinement level of the grid. After a relaxation phase of $80 \Myr$ we activate the SF and the feedback. After another $15$~Myr of evolution, the simulation reaches the maximum spatial resolution with all physical processes activated. We then let the system evolve for $\sim 2.4\Gyr$. 

\begin{figure}
	\centering
	\includegraphics[scale=0.26]{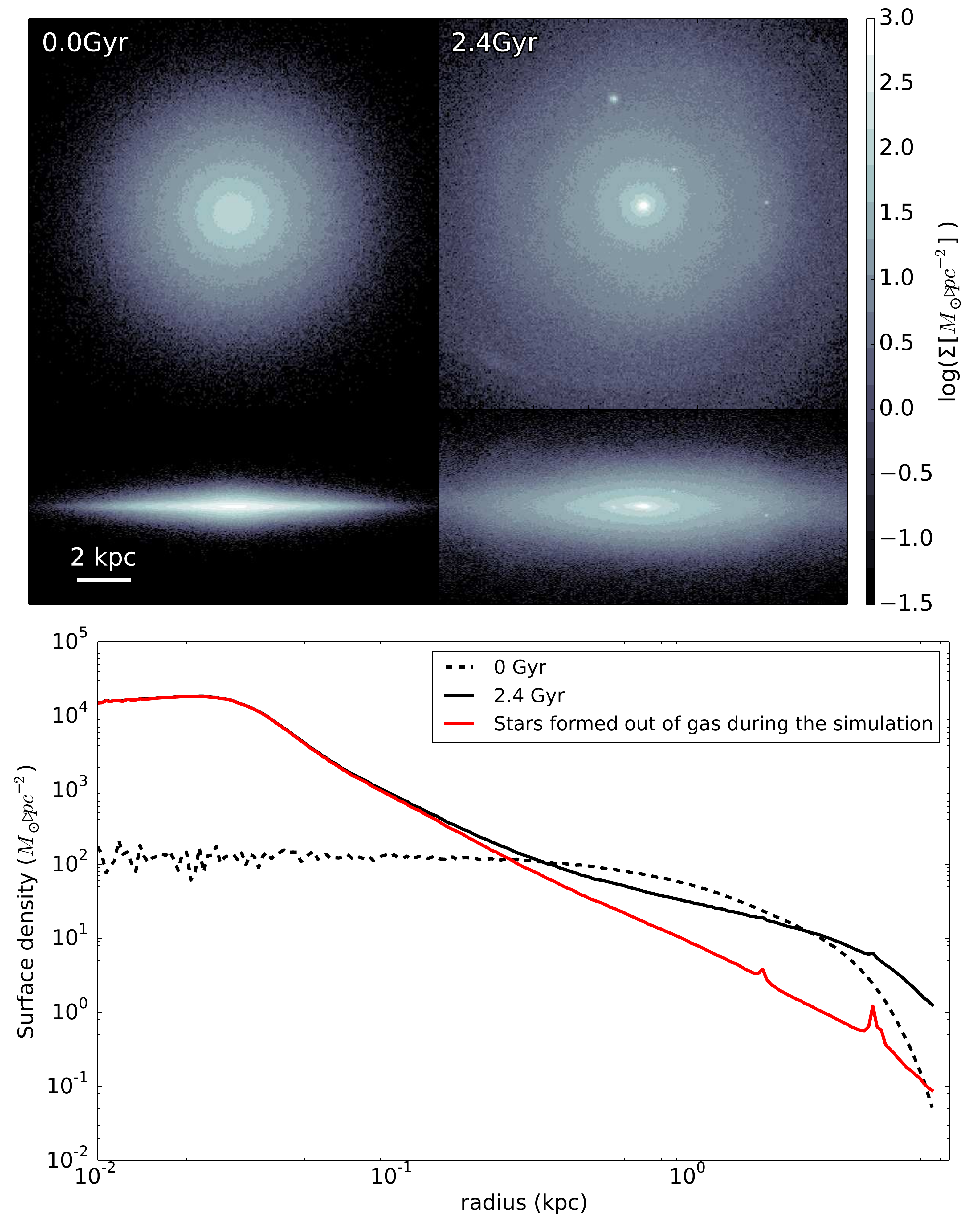}
    \caption{Top: Face-on and edge-on surface density maps of all stars at the beginning (left) and at the end (right) of the simulation. 
    Bottom: Radial profile of the surface density of the galaxy at the beginning (dashed) and the end (solid) of the simulation. The new stars (red) dominate the central hundred parsecs.}
    \label{fig:galaxy_allstars}
\end{figure}

At the end of the simulation, our galaxy has a stellar and gaseous mass of $1.5 \times 10^9 \Msun$ and $3.1 \times 10^8 \Msun$ respectively and a nuclear cluster has formed with a surface density of $2\times10^4$~\mpcc (see right panel of Fig.~\ref{fig:galaxy_allstars}).

We detect three smaller clusters orbiting around the nucleus with a period of a few hundreds Myr and orbital eccentricity between 0.3 and 0.6. The radial profile of the galactic surface density can be decomposed in three parts: the central region ($R < 200$~pc) which is dominated by new stars, a transition range for $0.2 < R [\mbox{kpc}] < 1$ and the outer part of the galaxy that exhibits an exponential profile with a scaling radius of $1.7$~kpc. 



\subsection{Numerical convergence}
\label{num_convg}

\begin{figure}
	\centering
	\includegraphics[scale=0.4]{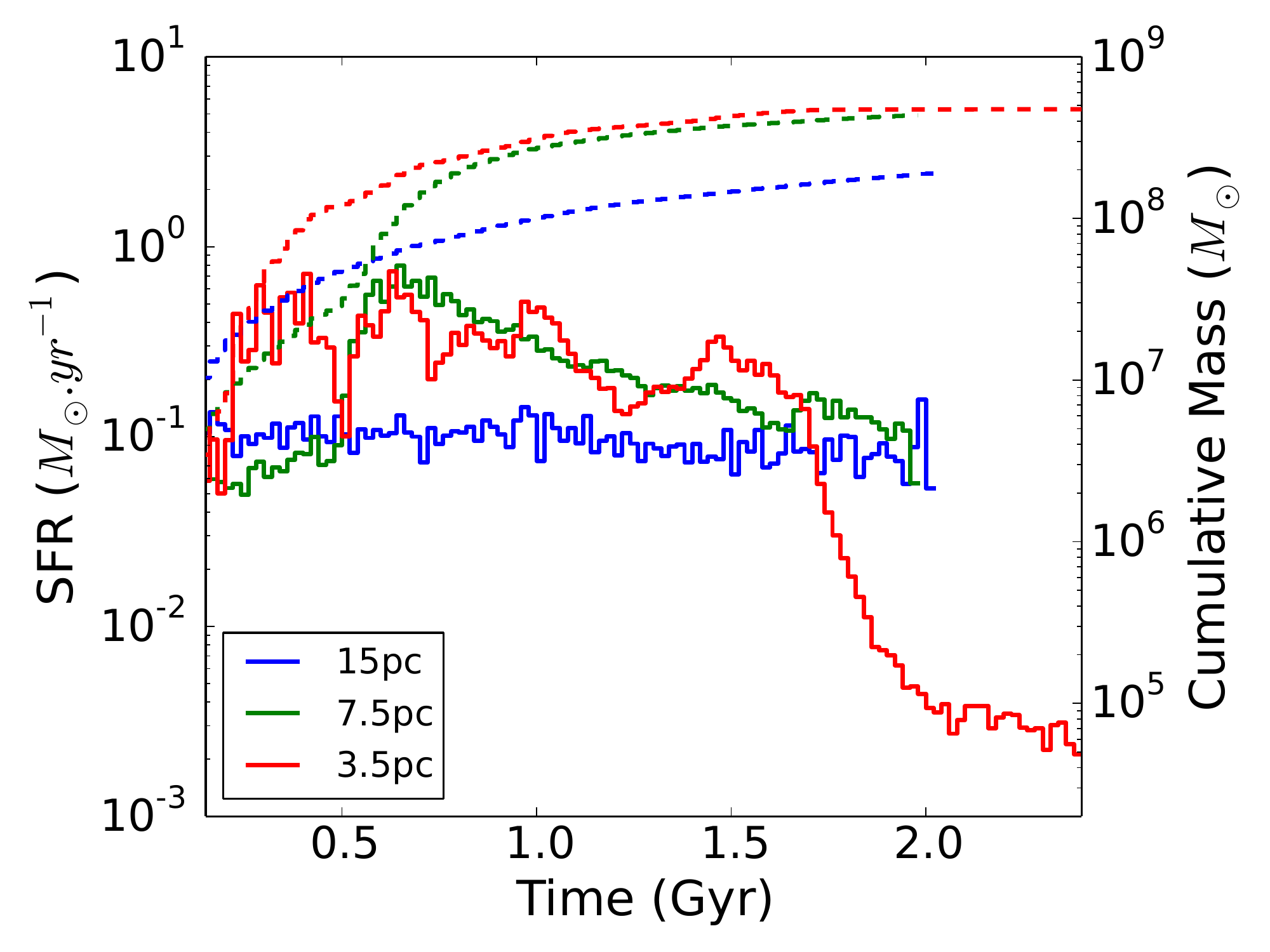}
   \caption{Cumulative Mass of new stars (dashed) and SFR (filled) for simulations at resolutions of $15 \pc$ (blue), $7.5 \pc$ (green) and $3.5 \pc$ (red). 
	    At t$\sim 1.7 \Gyr$, a merger between two massive clusters occurs, coinciding with a sharp drop of the SFR.}
  \label{fig:sfr}
\end{figure}

When increasing the resolution, we have access to denser regimes of gas, which thus potentially increases the SFR. This increase is however regulated by feedback. We tested the efficiency in providing numerical convergence by running two additional otherwise identical simulations with maximum resolutions of $15 \pc$ and $7.5 \pc$, respectively. Considering the complex evolution in the early stages of these gas-rich simulations, it is not relevant to compare the local details (star formation distribution, high frequency features, etc) of each simulation. Still, it is important to figure out if the global properties do converge. Figure~\ref{fig:sfr} thus shows that the SFR is quantitatively different between the 15 and 7.5~pc resolution simulations. The former has an almost constant SFR, while the latter shows a rapid increase within the first 500~Mr and a steady decrease hereafter. In that context, the 3.5~pc resolution simulation shows a very similar behaviour, even though the higher resolution allows to capture higher gas densities. This is confirmed by the fact that for the cumulative mass of new stars, convergence in the final stellar mass seems to occur between $7.5 \pc$ and $3.5 \pc$. In short, the simulations at $7.5\pc$ and $3.5\pc$ form about the same amount of stellar mass  ($4.8 \times 10^8 \Msun$) by the end of the simulation. In the rest of the paper, we thus focus on the simulation at the highest resolution, i.e. $3.5\pc$.



\subsection{Clusters detection}
\label{clst_find}
To detect star clusters, we use the friend-of-friend algorithm HOP \citep{Eisenstein1998}. With this method, clusters are defined as over-densities regions above a given threshold. Namely, a cluster is detected when the peak of the local stellar density exceeds 1.5\mpc. Two clusters are then merged if the saddle density between them is higher than 1\mpc. 
The clusters properties can be significantly affected by the choice of parameters in this algorithm. Lowering the densities would obviously result into a contamination from the background stars, while increasing it would lead to more compact (detected) clusters. We test that changing the detection parameters by a factor of two slightly affects the derived properties of the clusters, but does not alter the conclusions of the paper.



\begin{figure*}
	\centering
	\includegraphics[scale=0.55]{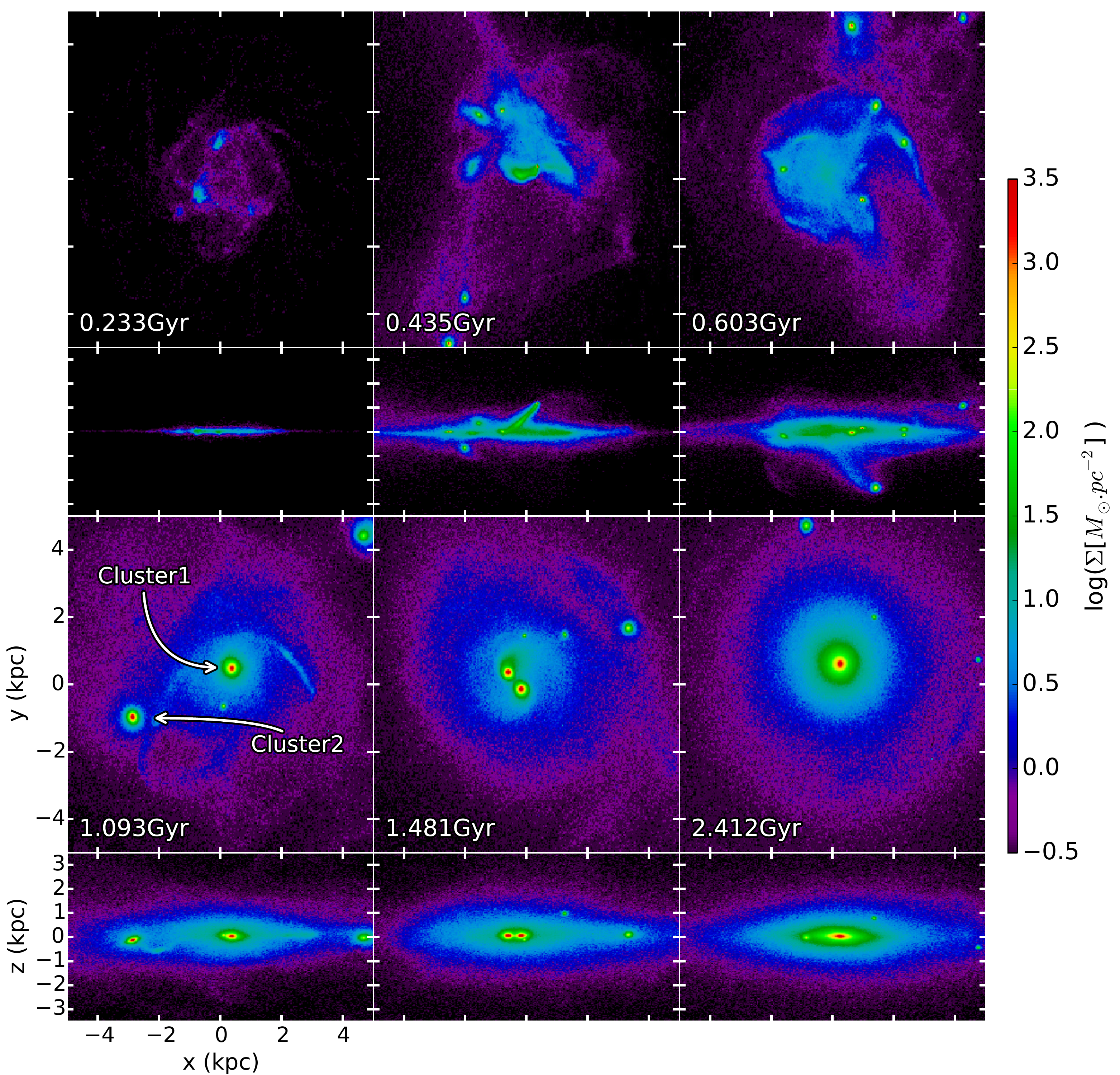}
    \caption{Surface density of stars that have been formed during the simulation. Cluster1 is the NC's seed. Cluster2 is the second most massive cluster in the simulation. It merges with Cluster1 at t=1.7 Gyr.}
    \label{fig:ns_all_global}
\end{figure*}

\section{Formation of NC}
\label{story_explanation}

Based on the simulation, we propose a new scenario for the formation of nuclear clusters. This \textgravedbl wet migration\textacutedbl~ scenario consists of two main phases: the formation, growth and migration of a massive cluster toward the centre of the galaxy during which the cluster retains part of its gas, followed by a potential merger with another cluster.



\subsection{Formation by migration}
\label{clst_form}

\begin{figure}
	\centering
	\includegraphics[scale=0.325]{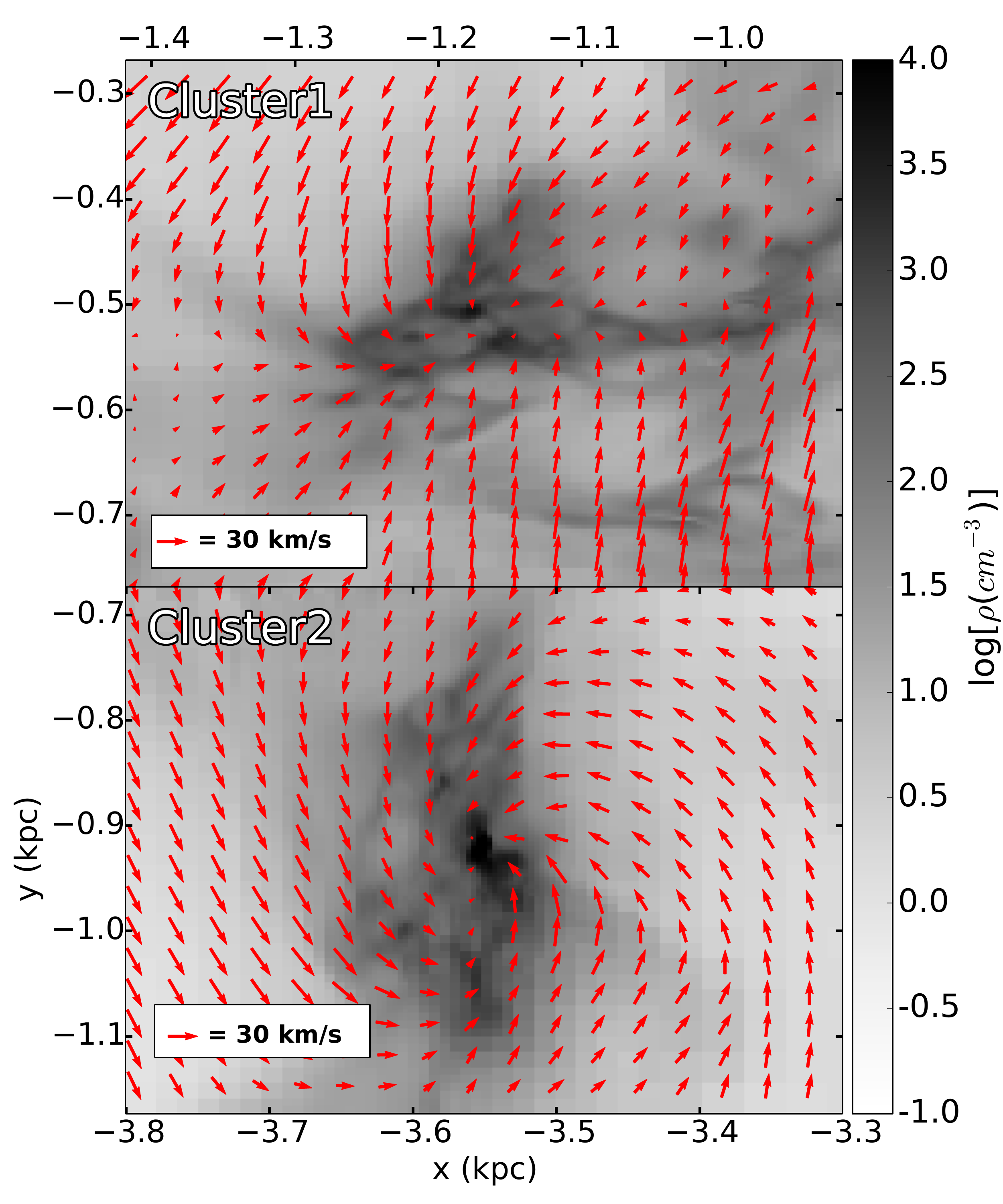}
   \caption{Maps of gas density at the earliest detection of the two most massive clusters in the galaxy. Cluster1 (top) forms the nuclear cluster by migration, while Cluster2 (bottom) merges later with the NC. The velocity field in the (x,y) disc plane is shown with 
   red arrows.}
  \label{fig:clst_formation}
\end{figure}

\begin{figure}
	\centering
	\includegraphics[scale=0.41]{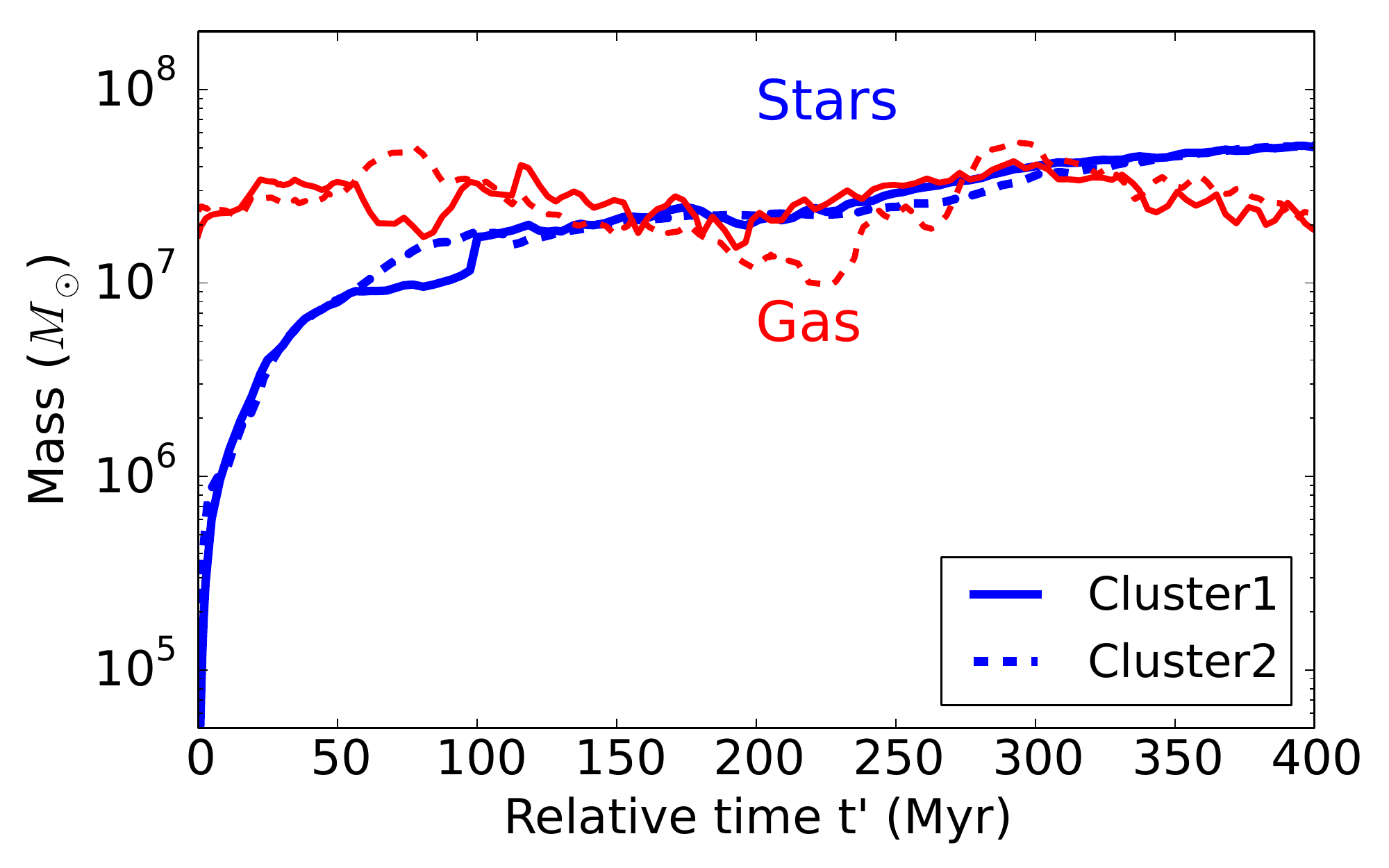}
   \caption{ In blue: stellar mass of Cluster1 (solid) and of the second most massive cluster Cluster2 (dashed) starting at their respective first detection. In red: gas mass within $200 \pc$ around the clusters. $t'=0$ corresponds to the respective earliest detection epoch of the clusters. At $t'=100 \Myr$, Cluster1 merges with another cluster which rapidly increases its mass.}
  \label{fig:clst_mass_growth}
\end{figure}

The cluster seed of our NC (named Cluster1) forms $1.1 \kpc$ away from the galactic centre,  at $t=562\Myr$.
At this stage, gas is still the major baryonic component of the galaxy disc, which has a rather irregular structure (see Fig.~\ref{fig:ns_all_global}).
A variety of clusters also form at the same epoch, with masses ranging from $10^5 \Msun$ to $10^7 \Msun$.
Cluster1 collapses out of a clump of $\sim 2\times10^7 \Msun$ ($\sim$0.8\% of the galactic gas mass, see top panel of Fig.~\ref{fig:clst_formation}). The initial cluster has a stellar mass of $2\times10^4 \Msun$, and
converging flows supply the cluster with gas (see the gas velocity field in Fig.~\ref{fig:clst_formation}). 
The gravitational potential of Cluster1 is deep enough to retain this reservoir, keeping 
a relatively constant mass of gas ($2-3\times10^7\Msun$) in its vicinity despite its stellar feedback. 
Sustained star formation makes Cluster1 steadily grow in mass (see the solid lines in Fig.~\ref{fig:clst_mass_growth}). Cluster1 also grows in size from $\sim12\pc$ to $30-40 \pc$.

\begin{figure}
	\centering
	\includegraphics[scale=0.41]{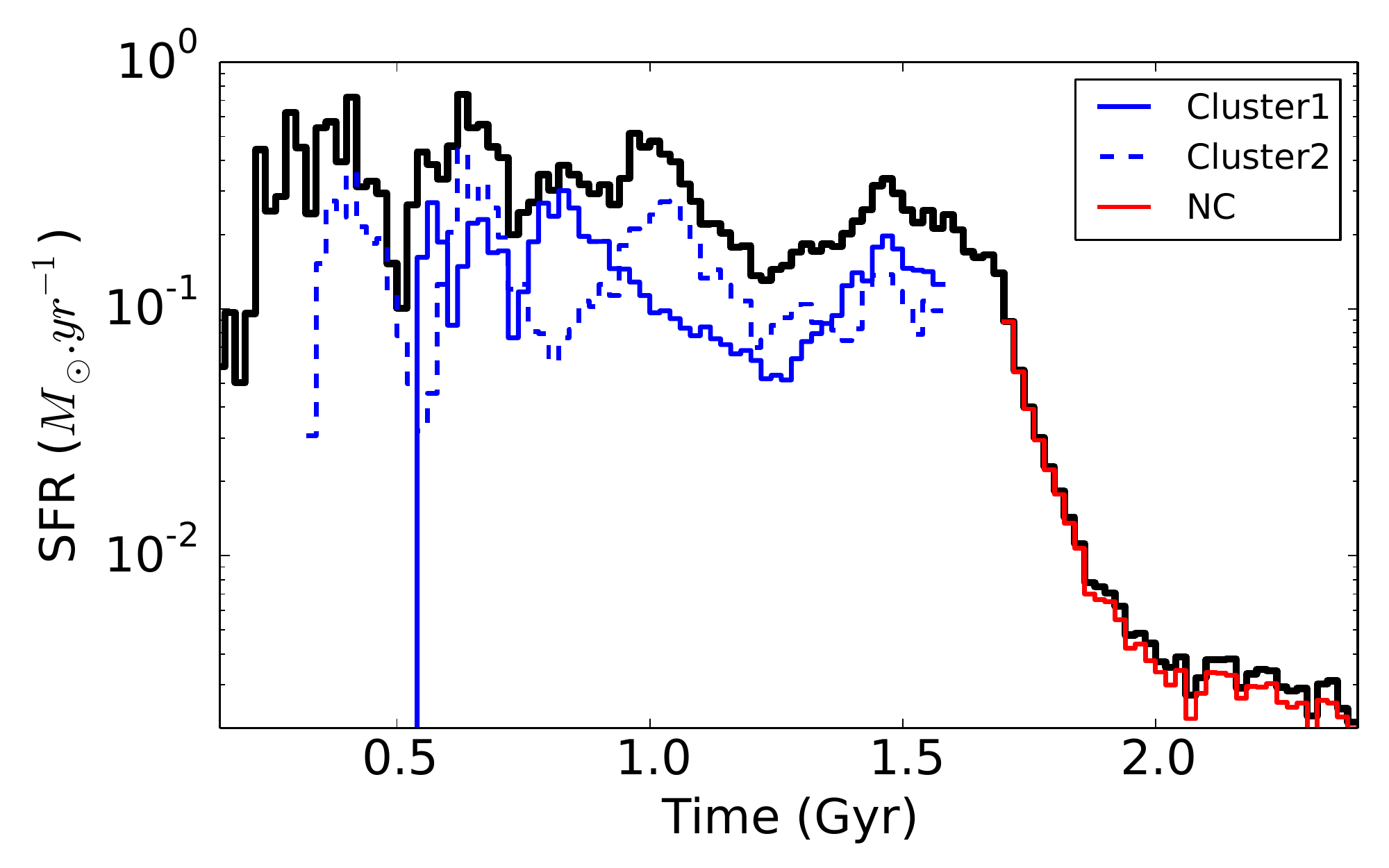}
   \caption{Contribution of Cluster1 (solid), Cluster2 (dashed) and their merger (NC, red) in the total SFR (black). The latter is dominated by Cluster1 and Cluster2 and by the NC in the end. Cluster1 and Cluster2 cannot be distinguished from each other after $t=1.6\Gyr$.}
  \label{fig:clst_sfr}
\end{figure}

We can split the growth of Cluster1 into two phases: 
\begin{enumerate}[1)]
 \item a rapid growth during the first $100 \Myr$. The gas dominates the mass budget within $200\pc$. 
 \item a slower growth in the following $800 \Myr$ during which the mass of Cluster1 dominates the environment over the gas reservoir.
\end{enumerate}
During the first phase, the amount of gas ($>10^7 \Msun$) remains higher or of the same order of magnitude than the mass of Cluster1 (see Fig.~\ref{fig:clst_mass_growth}). Variations of the gas reservoir mass have a strong impact on the mass growth rate of Cluster1: a decrease of the reservoir mass stops the growth (e.g. at $t'=60 \Myr$ where $t'$ is the relative time after the cluster formation) and its refilling accelerates it (e.g. at $t'=100 \Myr$). The refilling occurs both by local infall and during interactions with another dense cluster bringing its own gas. The decrease is mostly due to star formation and to SN blasts from the cluster itself or its neighbours. 
Fig.~\ref{fig:clst_sfr} shows that, since its formation, Cluster1 is one of the main contributors to the global SFR.

\begin{figure}
	\centering
	\includegraphics[scale=0.4]{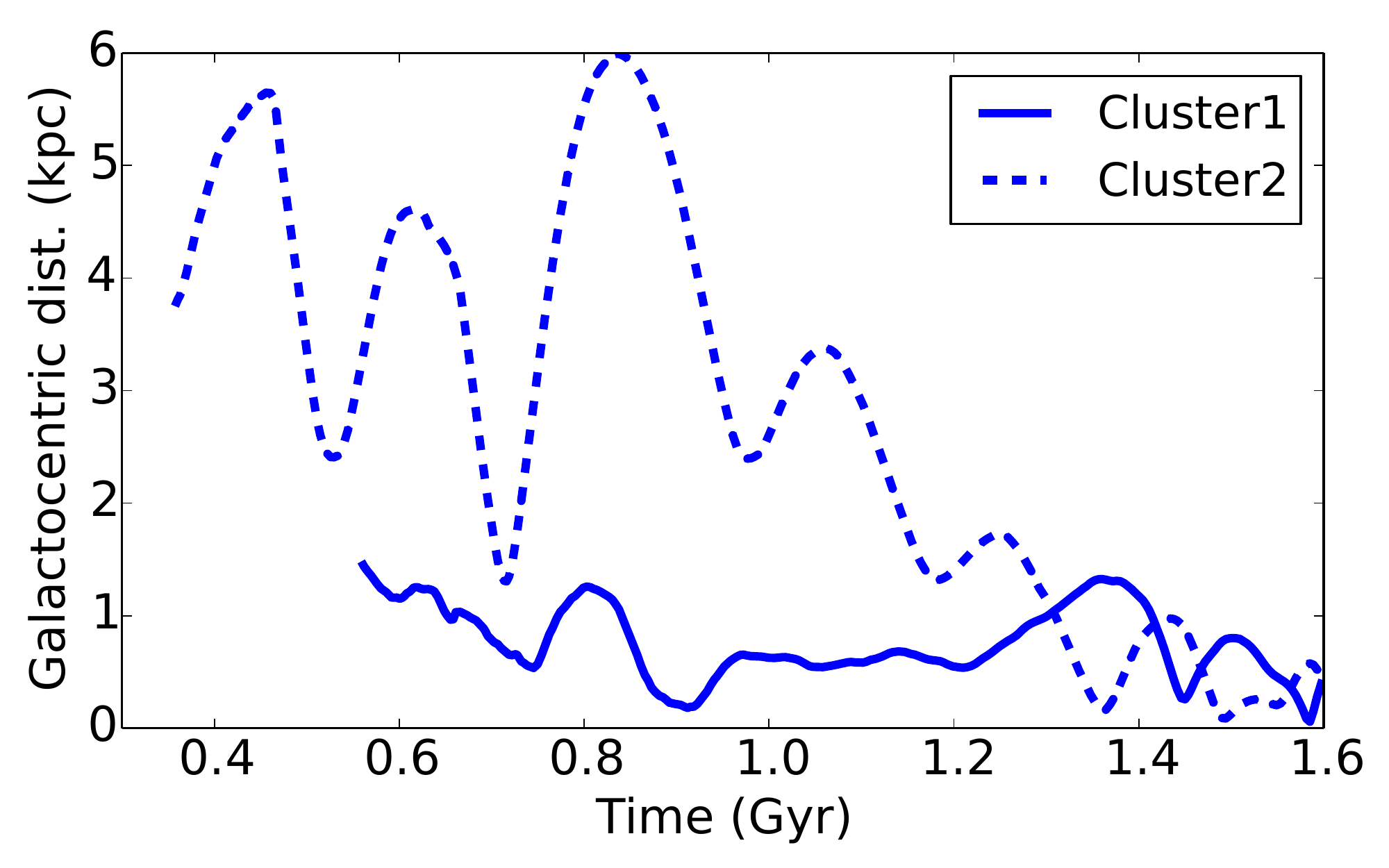}
   \caption{Galactocentric distance of Cluster1 (solid) and Cluster2 (dashed). The galactic centre is defined as the centre of mass of particles (stars + DM). }
  \label{fig:clst_migration}
\end{figure}

Fig.~\ref{fig:clst_migration} shows that Cluster1 migrates toward the centre relatively slowly. Indeed, it takes $350 \Myr$ to Cluster1 to cover a radial distance of $1.2 \kpc$ at $t<900\Myr$. Multiple interactions between Cluster1 and the surrounding structures slightly affects its orbits, and disturbs its migration towards the galactic centre.

\begin{figure}
	\centering
	\includegraphics[scale=0.375]{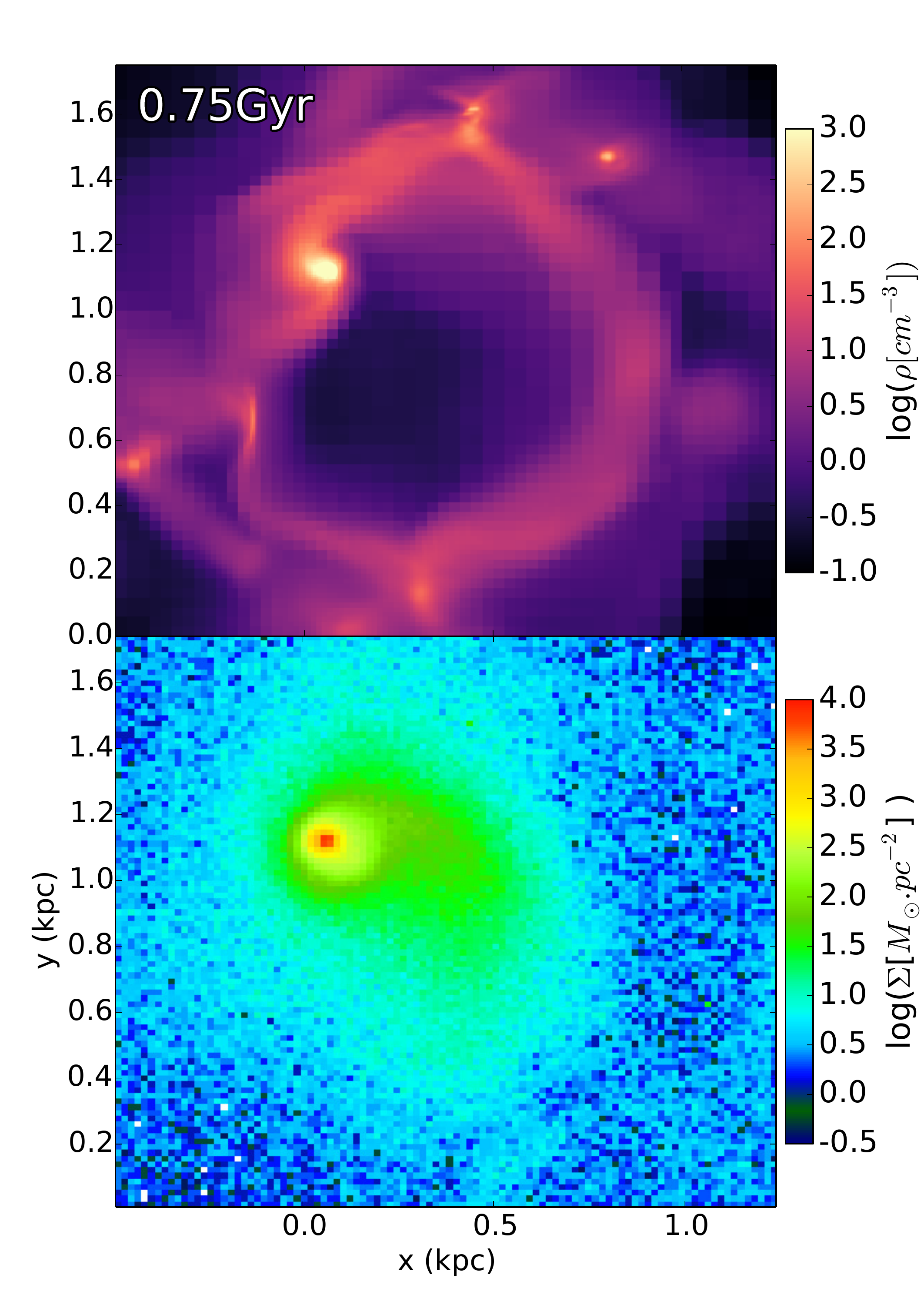}
   \caption{Maps of the gas (top) and new stars (bottom) densities. The shell from the supernova which explodes $5 \Myr$ before is visible on the gas density map.  The asymmetric extension of Cluster1 (on the right) is the combined result of its orbit and of the supernova blast.}
  \label{fig:clst_disturb}
\end{figure}

SNe also have an impact on the orbital evolution of the cluster. For example, at $t=730\Myr$ ($t'=170\Myr$ in relative time), Cluster1 experiences a burst of star formation (see Fig~\ref{fig:clst_sfr}). The newly formed stars slowly drift away from the remaining gas clump (due to asymmetric drift e.g. \citealt{Renaud2013}). $10 \Myr$ later, SNe feedback inject energy into the ISM, forming a bubble which is therefore off-centred with respect to the gas clump (see Fig.~\ref{fig:clst_disturb}). Since the gas represents a significant fraction of the local mass budget (52\% at that time for Cluster1, see Fig.~\ref{fig:clst_mass_growth}), the local gravitational potential is significantly altered when the gas is expelled. As a result, Cluster1 gets a velocity kick which increases its orbital eccentricity, and sends it away from the galactic centre (see Fig.~\ref{fig:clst_migration}). About $50 \Myr$ later, the cluster reaches its apocentre and moves back towards the centre, reaching this time a smaller galactocentric distance ($d=180\pc$ at $t=900 \Myr$). At that stage, the cluster represents 67\% of the galactic central (r<500pc) mass. It interacts with the stellar and gaseous material in the central region of the galaxy, which makes the galactic centre ill-defined. Nevertheless, the orbit of Cluster1 remains close to the centre of the global potential so that we can then consider Cluster1 as a NC. 

\subsection{NC-cluster merger}
\label{clst_merger}

Another massive ($4\times10^7 \Msun$) cluster (Cluster2) evolves alongside Cluster1. 
It forms in a different environment (see bottom panel of Fig.~\ref{fig:clst_formation}) in the external region of the galaxy ($d = 3.8 \kpc$, $t=360\Myr$) where the stellar and gas densities are much lower. 
The ISM around Cluster2 is slightly less turbulent than around Cluster1 (Mach number of 0.33 and 0.66, respectively, on a scale of $\sim 240 \pc$).
The early mass evolution of Cluster2 is similar to that of Cluster1 (see Fig.~\ref{fig:clst_mass_growth}). 
Figure~\ref{fig:clst_sfr} shows that Cluster2 is another important contributor to the overall SFR in the galaxy. 
We also note that the stellar mass dominates Cluster2 $300 \Myr$ after its formation, like Cluster1. Cluster1 and Cluster2 are thus 
initially in the same mass regime and share similar properties, while formed in rather different
environments.

Figure~\ref{fig:clst_migration} shows that after interactions with the sub-structures in the galactic disc ($t<950 \Myr$), 
Cluster2 loses angular momentum and progressively migrates towards the centre.
We estimate that the dynamical friction time is $\sim 1\Gyr$ (\citealt{Chandra1943}; \citealt{Mo2010}), which is consistent 
with the time Cluster2 takes to reach the central region of the galaxy. At $t=1.7\Gyr$, Cluster2 merges with the NC (initially Cluster1, which migrated earlier). The resulting stellar system has a half-mass radius of $\sim 35 \pc$ and a mass of $1.8\times10^8 \Msun$ 
(see bottom-right row of Fig.~\ref{fig:ns_all_global}). 
Because of the transfer of orbital momentum from Cluster2 to the stars of the merger, the resulting NC is flattened in the orbital plane of the interaction (which coincides with the plane of the galactic disc), with an axis ratio of 0.4 \footnote{We estimate the height and radius using iso-surface density contours of $10^3 \mpcc$ in its edge-on projection.}.

\begin{figure}
	\centering
	\includegraphics[scale=0.4]{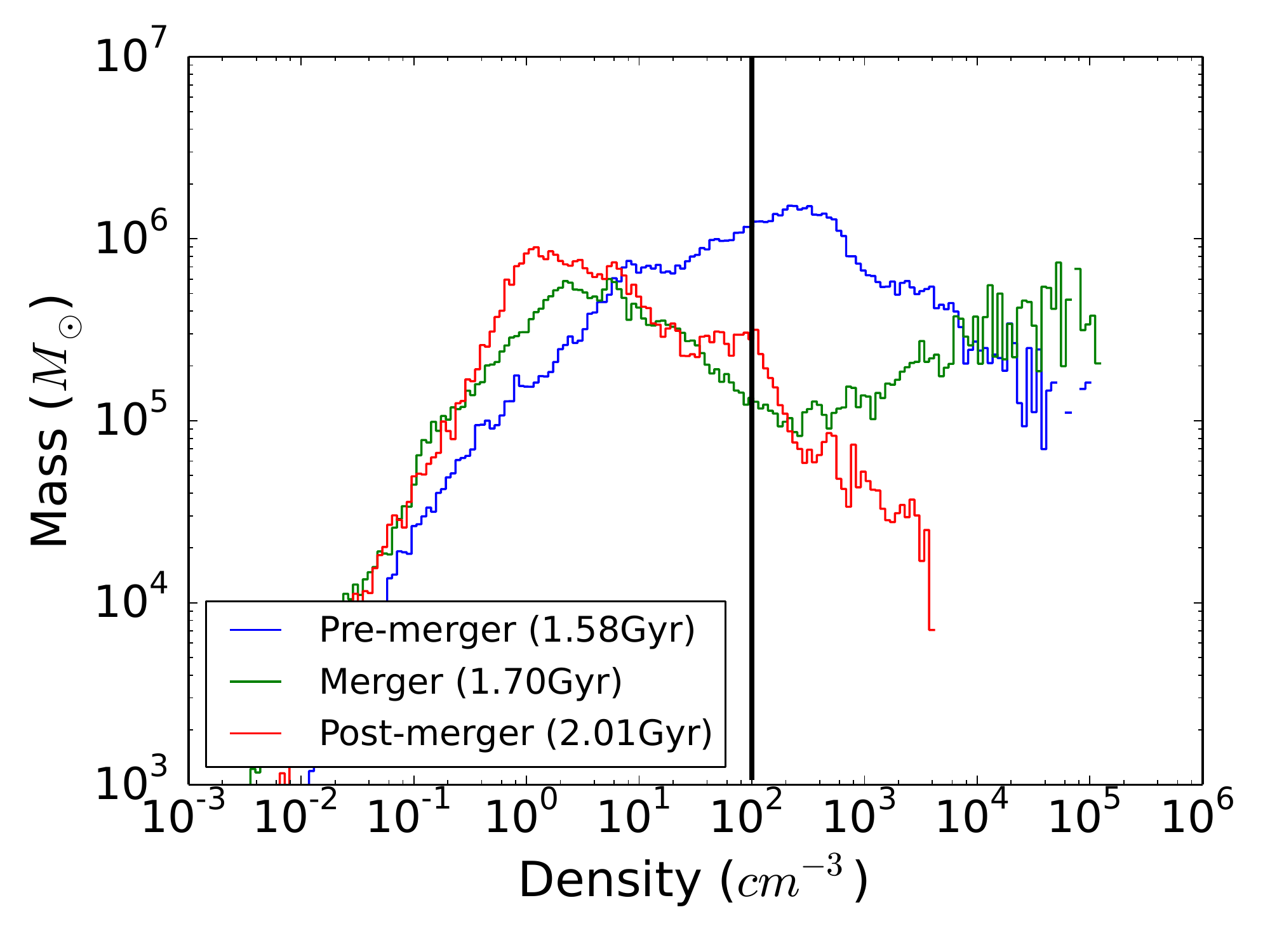}
   \caption{PDFs inside a $1\kpc\times1\kpc\times1\kpc$ centred on the NC or centre of mass of the system (Cluster1-Cluster2) before they merge. The vertical line represents the density threshold for the SF.} 
  \label{fig:pdfs_nc}
\end{figure}

After the merger, the SFR drops by almost two orders of magnitude (see Fig.~\ref{fig:clst_sfr}). Fig.~\ref{fig:pdfs_nc} shows the evolution of the gas density Probability Distribution Function (PDF) within the central kpc, during the merger phase.
Before the merger, the PDF yields a classical log-normal shape corresponding to supersonic ISM \citep{Vazquez1994}, and a power-law tail for $\rho \gtrsim 2000 \mathrm{cm^{-3}}$ indicating self-gravitating gas \citep{Elmegreen2011, Renaud2013}. The collision between the gas clouds around the NC and Cluster2 generates an excess of dense gas ($>10^4 \cc$), leading to a starburst localized in the central 25 pc. In the mean time, the tidal interaction strips gas from the outskirts of the clouds, thus depleting gas at intermediate density ($\sim100 \cc$). 
The dependence of star formation on $\rho^{3/2}$ implies that the depletion at intermediate densities approximately balances the central excess at high densities. Thus, despite the central mini starburst, the net SFR remains almost constant over 100 pc.
After the merger, the central star formation has consumed a large fraction of the dense gas, and the associated feedback disperses most of the gas left in this volume. This lack of dense gas reduces significantly the SFR to a few $10^{-3} \Msunyr$, thus almost quenching star formation in the NC.

\begin{figure}
	\centering
	\includegraphics[scale=0.4]{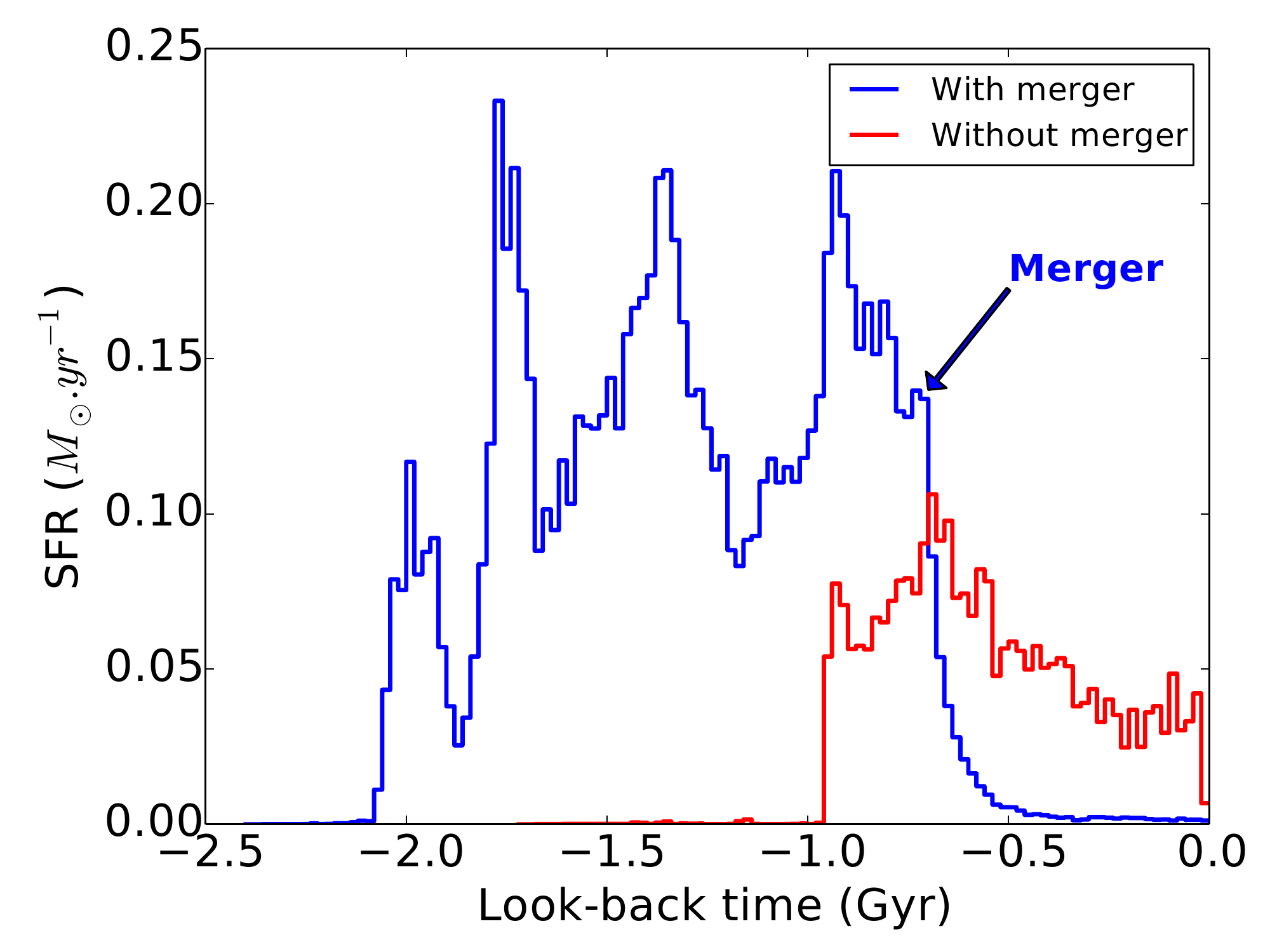}
   \caption{Star formation history $700\Myr$ after the formation of the final NC (t=0). Only the stars within a radius of 100pc centered on the NC are considered. When the NC experiences a merger (arrow at $t=-0.7\Gyr$), the stellar population of the NC is a mix of that of Cluster1 (formed at $t=-1.9\Gyr$) and that of Cluster2 (formed at $t=-2.1\Gyr$). Star formation is quenched after the event. When there is no merger, the stellar populations is only that of the NC progenitor (formed at $t=-0.9\Gyr$).} 
  \label{fig:nc_stelpop}
\end{figure}

\begin{figure*}
	\centering
	\includegraphics[scale=0.425]{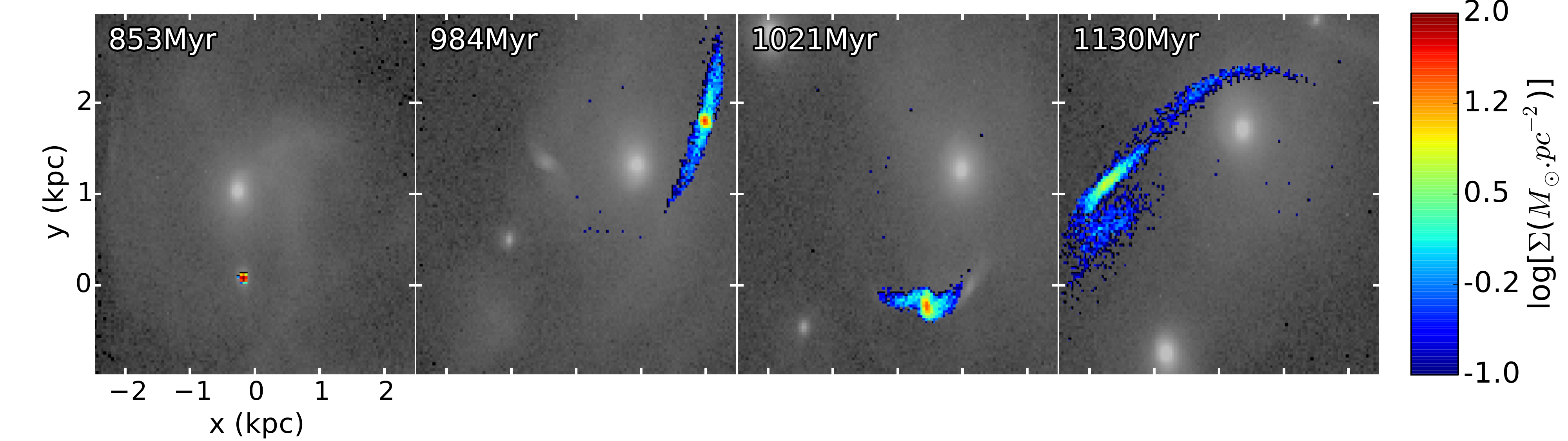}
    \caption{Disruption of a cluster through time. The galactic stellar background is shown in grey scale (Cluster1 is the closest to centre of the maps). The colour scale represents the surface density of the stars initially detected in the cluster (first panel). Interactions between clusters in the galaxy (mainly Cluster1 and Cluster2) generate tidal tails and eventually lead to the dissolution of the low density cluster. }
  \label{fig:cluster_disruption}
\end{figure*}

\subsection{Merger: a mandatory process?}
\label{clst_merger_manda}

To test the importance of the merging step in the formation scenario of the NC, we artificially remove the stars associated with Cluster2 from the simulation, before it interacts with cluster1 ($t=500\Myr$, i.e. when Cluster2 has formed about half of its final mass). This procedure is sufficient to prevent the further formation of a massive cluster, and does not alter the large scale dynamics of the rest of the galaxy.

In this alternative simulation, a cluster similar to Cluster1 still forms at $t=800\Myr$ and reaches the centre in about the same amount of time, namely $300\Myr$. The NC forms as described in Section~\ref{clst_form}. We then let the NC evolve for $700\Myr$ ($t=1.7\Gyr$).
However, the absence of another massive cluster being able to merge with the NC voids the second step of our scenario. All the effects associated with the merger phase (recall Section~\ref{clst_merger}) are thus missing in the further growth of the NC.
 Namely,
\begin{itemize}
\item the depletion of the dense gas reservoir does not occur and the NC continuously forms stars. This affects the star formation history of the NC as shown in Fig.~\ref{fig:nc_stelpop}. In the merger scenario, both NC cluster progenitors form stars during their entire lifetimes, until star formation gets quenched at the time of the collision. This leads to the mixing of stellar populations with different ages, and the lack of a young population.
\item the angular momentum re-distribution noted during the merger does not happen and the NC maintains an almost spherical morphology (axis ratio of 0.8), as opposed to the flattened shape visible in Fig.~\ref{fig:ns_all_global}. 
\item Without merger, there is no increase of the angular momentum and the resulting NC exhibits a lower amplitude rotation than in the case of a merged system: the difference in angular momentum is approximately of a factor of 10.
\item the resulting NC is less massive but has a similar size ($5\times10^7\Msun$ and $40\pc$ in our cases) without the merger step. 
\end{itemize}
Note that the first three points could be used as observational diagnostics to establish the formation scenario of real NCs.

This demonstrates that the merger step is not mandatory for the formation of the NC, but can significantly alter the properties of the NC when it takes place.



\section{Cluster populations}
\label{clst_small}



\subsection{Cluster disruption}
\label{clst_disr}

In our fiducial simulation, Cluster1 and cluster2 represents ~15\% of the new stars of the disc, and set the dynamics of their surroundings. The rest of the star cluster population thus experiences several interactions with Cluster1, Cluster2 and the NC, and some get disrupted by tidal forces. Signatures of tidal disruptions are visible throughout the simulation (see e.g., bottom-left panel of Fig.~\ref{fig:ns_all_global}). 

One example of this disruption process is shown in Fig.~\ref{fig:cluster_disruption}, where we monitor the stars of one cluster during about 300~Myr, until its complete destruction by tidal forces. At $t= 853 \Myr$ (first panel of Fig.~\ref{fig:cluster_disruption}), a bound cluster is detected $1 \kpc$ away from Cluster1. The tidal interaction between Cluster1 and this $\sim10^5\Msun$ cluster induces tidal tails (second panel of Fig.~\ref{fig:cluster_disruption}). Subsequent interactions, including one with the approaching Cluster2, accelerate the disruption, finally leading to complete dissolution. The tidal features can still be detected as elongated over-densities for another $250\Myr$ after the dissolution, until the surface density contrast with the background becomes too low. 
This situation is similar for other less dense clusters. This shows the key role of massive clusters such as Cluster1 and Cluster2 in the evolution of the cluster population as a whole, accelerating the disruption of the most fragile objects.



\subsection{Surviving clusters}

\begin{figure}
	\centering
	\includegraphics[scale=0.4]{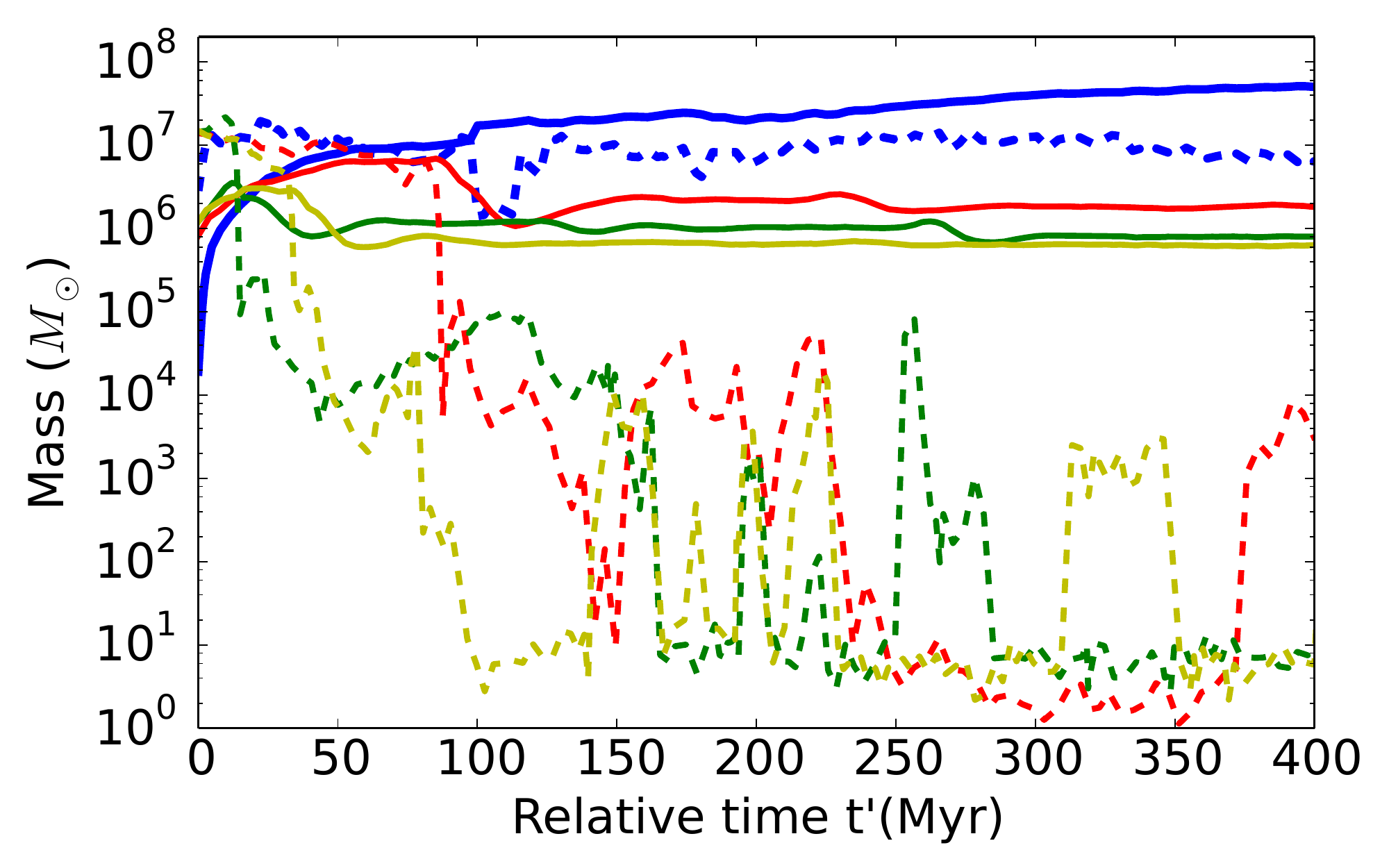}
   \caption{Evolution of the stellar (solid) and gas (dashed) mass of the clusters surviving the presence of the NC, compared to that of Cluster1 (blue). We measure the gas mass in a cube of $50\pc$ centred on the cluster. Each color corresponds to one cluster (A in red, B in green and C in yellow). As in Fig.~\ref{fig:clst_mass_growth}, the time is relative, with $t'=0$ marking the earliest detection of each cluster.}
  \label{fig:clst_mass_growth_cluster}
\end{figure}

\begin{figure}
	\centering
	\includegraphics[scale=0.4]{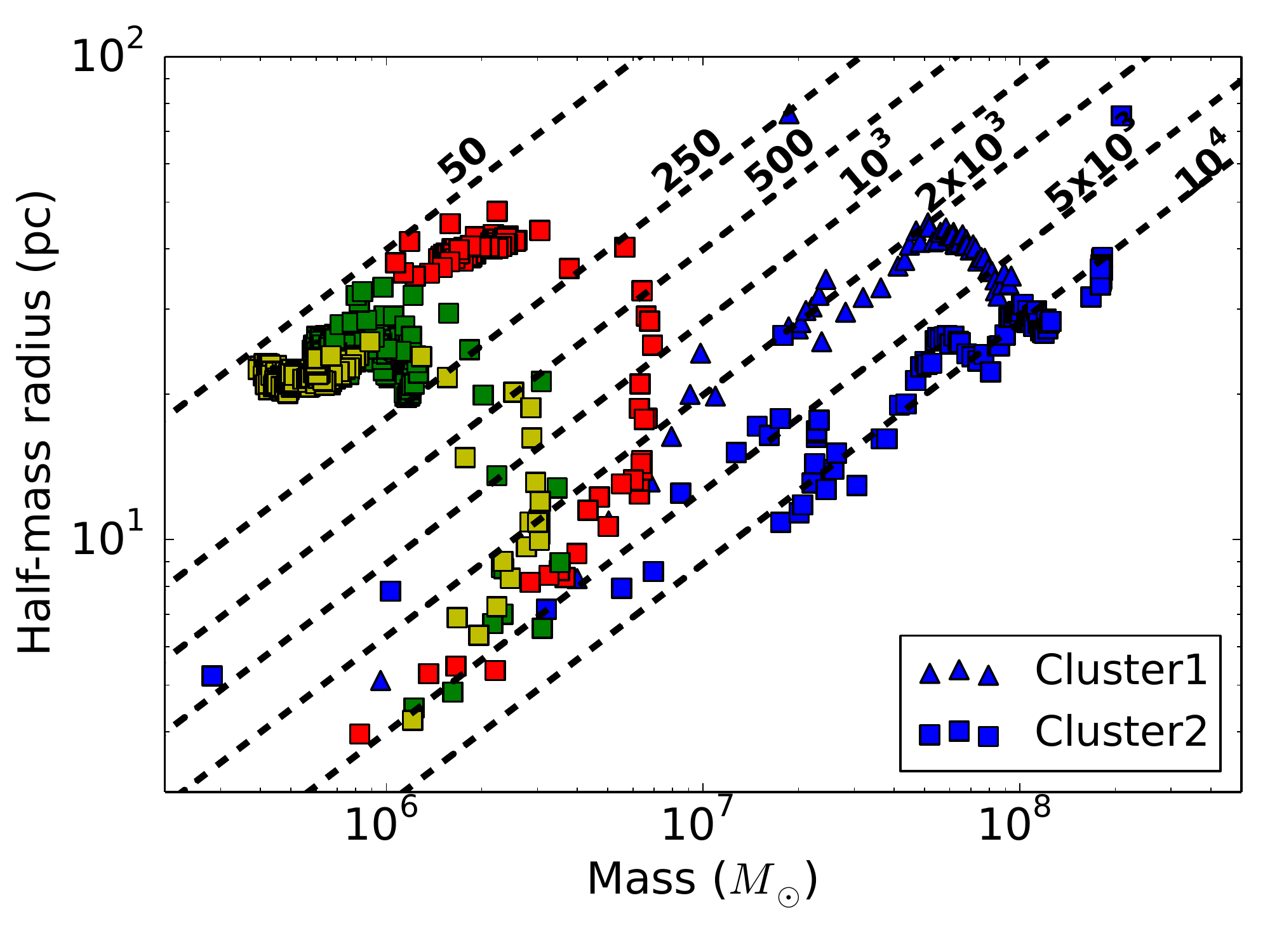}
   \caption{Evolution of size and mass of all clusters detected at the end of the simulation. Colours are as in Fig.~\ref{fig:clst_mass_growth_cluster}. The dashed lines follow constant surface density values in $\mpcc$.}
  \label{fig:clst_mass_size}
\end{figure}

\begin{figure}
	\centering
	\includegraphics[scale=0.4]{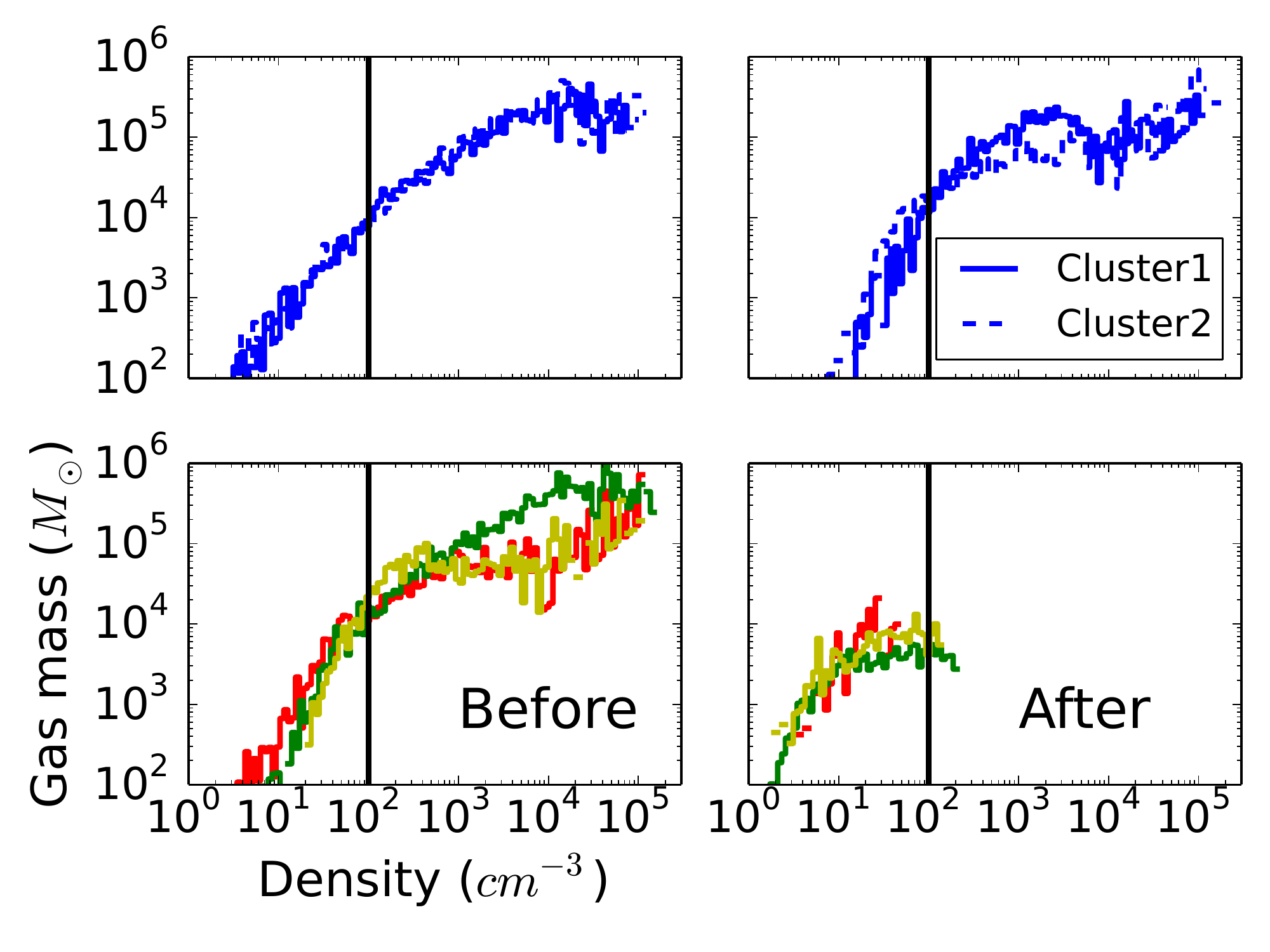}
   \caption{Gas density PDFs in regions of $50\pc$ centred on the clusters $5\Myr$ before (left column) and after (right column) the removal of gas by SN-blasts. The blue curve shows the evolution for Cluster1 and Cluster2 for reference. The vertical lines show the density threshold above which gas can be converted into stars.}
  \label{fig:clst_pdfs_blast}
\end{figure}

As illustrated in Fig.~\ref{fig:ns_all_global}, some clusters survive the disruptive presence of the NC.
We detect three of these clusters (named A, B and C) keeping a constant mass for most of the simulation ($\sim10^6\Msun$, see Fig.~\ref{fig:clst_mass_growth_cluster} and Fig.~\ref{fig:clst_mass_size}).
However, their mass evolutions strongly differ from that of Cluster1 and Cluster2.
Their growth phase only lasts about 10-40~Myr. This star formation activity leads to a rapid 
injection of supernova energy into the ISM, but their lower density is not enough to retain the feedback winds, which thus depleting the gas reservoir mass by one to three orders of magnitude (in mass). 
Figure~\ref{fig:clst_pdfs_blast} illustrates this by showing the evolution of the gas density PDFs in the regions of the clusters. 

For clusters A, B and C, stellar feedback happens to truncate the PDFs close to the density threshold associated with star formation, thus preventing further star formation. The rapid gas removal by stellar feedback in clusters A, B and C has a significant impact on the
local gravitational potential. The least bound stars are then ejected from the clusters \citep[]{Hills1980,Boily2003}. This lowers the clusters masses by a factor 2 to 7 and their surface densities by one order of magnitude (see Fig.~\ref{fig:clst_mass_size}), which then remain roughly constant until the end of the simulation. The mass of the gas reservoirs shows fluctuations over time. A sharp increase of the gas mass can lead to an increase of the clusters mass for a short period. This is for example the case for cluster B at $t'=250\Myr$ in Fig.~\ref{fig:clst_mass_growth_cluster}. The stellar mass of the cluster then decreases as the least bound stars are ejected from the cluster.

The main difference between NC progenitors and the rest of the cluster population is then their ability to retain a significant fraction of their stellar mass. In Cluster1 and Cluster2, the injected feedback energy is not high enough to significantly alter the existing gas reservoir. By keeping a dense gas reservoir, they can further form stars and become even more massive and resistant to subsequent tidal disruptions induced along their orbits. Altogether, these points indicate that the low density clusters have a much lower probability to survive and become seeds for a NC, in contrast with e.g., Cluster1 (see Section~\ref{clst_form}).

The dynamical friction time of clusters A, B and C is much longer, of the order of tenths of $\Gyr$, since the dynamical friction time is inversely proportional to the cluster mass. Therefore they cannot contribute to the building of the NC through mergers within several Gyr, unlike Cluster2 (see Section~\ref{clst_merger}).




\section{Discussion \& Conclusion}
\label{discussion}

Using hydrodynamical simulations of an isolated gas-rich dwarf galaxy, we propose a \textgravedbl wet migration\textacutedbl~scenario for the formation of nuclear clusters. The main steps are (see also Fig.~\ref{fig:scenario_schm}):
\begin{itemize}
\item A population of star clusters forms across the galactic disc.
\item Clusters dense enough to retain a gas reservoir around them maintain a star formation activity for a few 100 Myr, which steadily increases their masses.
\item These clusters loose orbital energy through dynamical friction and interactions with the rest of the disc and migrates to the centre to form a nuclear cluster.
\item The NC eventually experiences (wet) mergers with other dense clusters, increasing its mass and quenching its star formation activity.
\end{itemize}
The last step is not mandatory for the formation of the NC but strongly affect its properties (mass, shape, star formation history), as discussed in Section \ref{clst_merger_manda}.

The other star clusters in the galaxy have lower initial densities, which affects their early evolution and tells them apart from the NC progenitors. They are either tidally disrupted by the central structures (including the NC itself) or have high orbital angular momentum which prevents them to interact with the NC and participate to its build-up.

\begin{figure}
   \centering
   \includegraphics[scale=0.5]{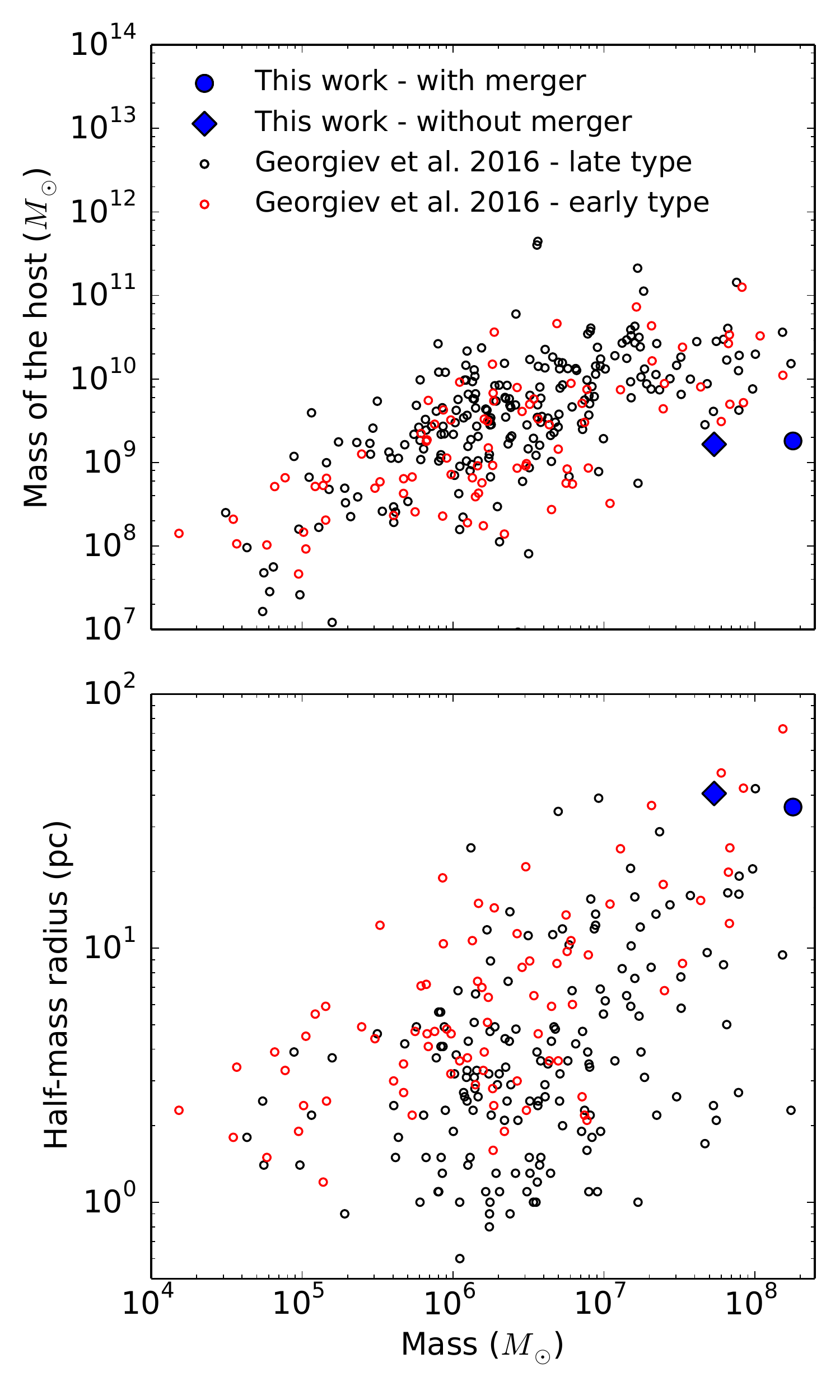}
  \caption{Position of the nuclear clusters formed in our simulation on the galaxy mass - observed cluster mass scaling relation (top) and in a size-mass diagram (bottom).}
 \label{fig:size_mass_diagram}
\end{figure}

By comparing the properties of the NC modeled with that of the observed population, Fig.~\ref{fig:size_mass_diagram} shows that our simulation is in line with the observed scaling relations \citep[e.g.][]{Georgiev2016}. Our NC lies in the high mass and size regime (40~pc and $5\times10^7 \Msun$ without merger, and 35~pc and $1.8 \times 10^8 \Msun$ with merger). Although well within the dispersion of observational data, NCs in this mass range would be preferentially detected in slightly more massive galaxies. However, galaxies with different masses are likely to play different roles on the formation process of their NCs, as underlined by previous works. Observations by \citet{denBrok2014} favour the migration scenario in the low-mass regime ($\lesssim 10^{9-10}\Msun$, see also the theoretical confirmation by \citealt{Arcasedda2014}). The relative important of in-situ star formation increases with galactic mass, as showed by \citet{Antonini2015}, suggesting that massive galaxies are more prone to drive gas flows toward the NC and fuel in-situ star formation than their low-mass counterparts. 

Such gas flows are related to kpc-scale dynamics of the galactic disc, in particular the presence of substructures. For instance, torques from bars are well-know to drive gas infall towards the galactic centre \citep{1979ApJ...233...67R,Athanassoula1992,Garcia2014,Emsellem2015}. 
This process would then supply the nuclear cluster with gas and maintain its star formation activity over long timescales. 
Ongoing star formation would then occur preferentially in the plane of the galactic disc \citep{Seth2006, Boker2010, Feldmeier2015}, thus leading to a flattened NC. A similar morphology is predicted by our model in the case of a cluster merger. However in our case, the merger quenches star formation. Therefore, the absence of young stars in a flattened NC favours our merger scenario, while a young population denotes in-situ formation.

We also note that spiral arms would lead to star cluster formation providing more candidates for dry or wet mergers with the NC. It is however not clear whether these potential NC progenitors would survive the radial migration through spirals and bars \citep{Fujii2012}. Probing these processes would require to model galaxies of different masses and disc stabilities over several rotation periods to allow for the formation and evolution of substructures.

Accounting for the cosmological context would also be key for replenishing the gas reservoir with low metallicity gas (through cold gas accretion), and/or triggering the formation and destruction of spirals and bars \citep[see e.g.][]{Kraljic2012}. Over  such long timescales, and particularly in the redshift range considered here ($z \sim 2 -3$), it is likely that a dwarf galaxy like that modeled here would experience interactions with its environment, either with other galactic systems or with the inter-galactic medium. Depending on the state of the cluster (growing seed or fully formed NC), we expect the outcome of these interactions to vary. On one hand, the perturbations will likely disturb the orbit of the seed. On the other hand, the dwarf may suffer from tidal stripping and its NC, if already formed, may become an Ultra Compact Dwarf galaxy (see e.g. \citet{Pfeffer2013} or \citet{Norris2015}. The nature of the perturbation is also to be considered. A dwarf within a cluster-like environment will likely experience processes such as gas stripping or ram pressure. The resulting impact on the gas content can thus be significant, with either boosting or slowing down the growth of the NC progenitors. Along the same lines, mergers would possibly induce dramatic changes both in the morphology and star formation history of the galaxies. The migration and growth of an NC seed should be examined further in such contexts.

Such timelapses become comparable to the relaxation timescales of typical NCs ($\sim 10^9 \yr$, although the massive ends of the population, including our case yield much longer timescales $\sim 10^{11} \yr$, see also \citealt{Seth2006}). Following the co-evolution of the NC and its host over such long periods would then require to consider collisional processes to properly treat the internal physics. Among other internal mechanisms, a full treatment of stellar evolution would provide insights on the formation of stellar mass black holes in the NC and its progenitor clusters. Then, the possible merger step in our scenario would represent an important channel in the formation of intermediate and possibly supermassive black holes in galactic centres \citep{Seth2006, Antonini2015}, potentially followed by an active phase of the galactic nucleus.

\section*{Acknowledgements}

We wish to thank the anonymous referee for the helpful comments and suggestions that improved the paper.
This research was supported by the DFG cluster of excellence 'Origin and Structure of the Universe' (www.universe-cluster.de).
We acknowledge the support by the DFG Cluster of Excellence "Origin and Structure of the Universe". 
The simulations have been carried out on the computing facilities of the Computational Centre for Particle and Astrophysics (C2PAP). 
FR acknowledges support from the European Research Council through grant ERC-StG-335936.



\bibliographystyle{mnras}
\nocite{Georgiev2016} 








\bsp	
\label{lastpage}
\end{document}